\documentclass[preprint,aps]{revtex4}

\usepackage{graphicx}
\usepackage{dcolumn}
\usepackage{bm}

\begin{document}
\title{A novel directional coupler utilizing a left-handed material}
\author{Sanshui Xiao$^1$, Linfang Shen$^1$ and Sailing He$^{1,2,*}$}
\address{$^1$Centre for Optical and Electromagnetic Research\\
State Key Laboratory for Modern Optical Instrumentation\\
Joint Laboratory of Optical Communications of Zhejiang University, \\
Zhejiang University, Yu-Quan, Hangzhou,  310027, P. R. China\\
$^2$Division of Electromagnetic Theory, Alfven Laboratory\\
Royal Institute of Technology, S-100 44 Stockholm, Sweden\\
Corresponding author: \underline{sailing@kth.se}}
\date{\today}

\begin{abstract}
A novel directional coupler with a left-handed material (LHM)
layer between two single-mode waveguides of usual material is
introduced. The coupling system is analyzed with the supermode
theory. It is shown that such a LHM layer of finite length can
shorten significantly the coupling length for the two
single-mode waveguides. A LHM layer with two slowly tapered ends is used to avoid the reflection loss at the ends.

\end{abstract}

\maketitle




\section{INTRODUCTION}
Recently, a new type of electromagnetic materials which have
simultaneously a negative electric permittivity and a negative
magnetic permeability have generated a great interest
\cite{ves,shelby,pen2}. Such a material is often called a
left-handed material (LHM) since the electric field, the magnetic
field and the wave vector of an electromagnetic wave propagating
in it obey the left-hand rule (instead of the right-hand rule for
usual materials). In 1968, Veselago theoretically investigated
such a material and predicted some extraordinary properties such
as the negative refraction, Doppler shift and Cerenkov radiation,
etc \cite{ves}. Because of the unavailability of LHMs at that
time, his idea did not attract much attention until very recently.
Smith {\it et al.} have demonstrated experimentally recently the
phenomenon of negative refraction \cite{shelby}. Isotropic LHMs
have also been introduced (see e.g. \cite{iso}). Pendry predicted
that a LHM slab can behave like a perfect lens since the LHM slab
can amplify evanescent waves \cite{pen2}. In this letter, we
propose a  novel directional coupler  with a LHM layer between two
straight single-mode waveguides of usual material (i.e.,
right--handed material (RHM)).

A directional coupler is a fundamental planar lightwave circuit
(PLC) performing the functions of splitting and combining.
Combined with optical delay lines, directional couplers can be
used to construct Mach-Zehnder interferometers (MZIs),
interleavers  and wavelength division multiplexers/demultiplexers
(WDM).  The two parallel single-mode (SM) waveguides in the
central coupling region of a conventional directional coupler (of
PLC type) are usually very close to each other in order to reduce
the coupling length. The separation between the two SM waveguides
increases gradually in the input and output waveguide regions in
order to connect the directional coupler to fibers or two {\em
decoupled} (well-separated) SM waveguides (to construct e.g. a
MZI). Thus, bent waveguides, which introduce some bending loss and
increase the size of the device, have to be used in the transition
region (see e.g. \cite{bend}). In this letter, we show that by
putting a LHM layer of finite length in the middle the coupling
length for two parallel SM waveguides can be shortened
significantly (this can make the device much shorter; it is
desirable in a future network that many compact components can be
put in an integrated chip).  In our novel directional coupler, the
two parallel SM waveguides can be well-separated from each other
and thus there is no need to use bent waveguides (this is another
advantage of the present novel directional coupler).

\section{FORMULAS FOR CALCULATING THE COUPLING LENGTH}

\begin{figure}
\includegraphics[width=6in]{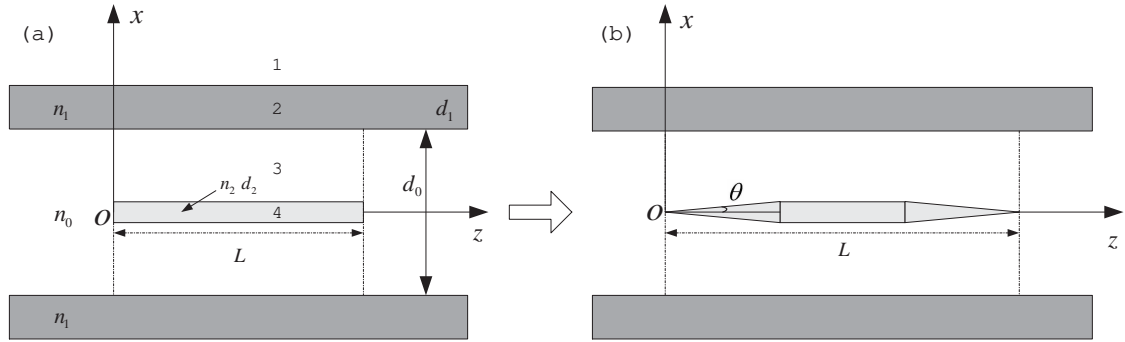}
\caption{\label{coupler}The schematic structure for the novel
directional coupler. (a) case when the ends of the LHM layer are
not tapered. (b) case when the LHM layer is tapered slowly at the
two ends.}
\end{figure}

In the design and analysis of directional couplers, the supermode
solution \cite{gne} has been widely used to calculate the coupling
length of two parallel waveguides. Compared to the coupled mode
theory, the supermode solution method is more accurate and is
still valid even the separation between the two waveguides are
quite small \cite{coup1}. Here we use the supermode solution
method to study the novel directional coupler shown in Fig.
\ref{coupler}. For a PLC, the effective index method \cite{eff}
can simplify the analysis greatly through converting the
three-dimensional model to a two-dimensional model. To increase
the coupling between the two SM waveguides, a left-handed material
(LHM) layer is placed between them. The refractive indices of the
core and the cladding for the SM waveguides are denoted by $n_1$
and $n_0$, respectively. The two SM RHM waveguides have the same
width $d_1$ and the separation between them is $d_0$. The negative
refractive index and width of the LHM layer are denoted by $n_2
(<0)$ and $d_2$, respectively. The propagation constants of the
two associated supermodes are denoted by $\beta_s$ (for the
symmetric supermode) and $\beta_a$ (for the antisymmetric
supermode). The region between the two vertical dash-dotted lines
in Fig. 1 indicates the strong coupling region.

Since the structure is two-dimensional, the electromagnetic fields
can be decomposed into the $E$- and $H$-polarized modes. Let us
consider the case of $E$-polarization (the $H$-polarization case
can be treated in a completely analogous way). Since the two
supermodes of this system is either symmetric or antisymmetric
about the central line $x=0$, we only need to study the field
distribution in the area $x>0$. The electric field of a supermode
in the four regions (see Fig. \ref{coupler} (a))  of the central
part (with strong coupling) can be written in,
\begin{eqnarray}
E_1 (x) = A\exp [ - p(x - (d_0 /2 + d_1))]\exp(i\beta z),  \qquad \mbox{in region 1},  \\ \label{ex1}
E_2 (x) = [B\cos (hx) + C\sin (hx)]\exp(i\beta z),  \qquad \mbox{in region 2}, \\ \label{ex2}
E _3(x) = [D\exp ( - px) + E\exp (px)]\exp(i\beta z), \qquad \mbox{in region 3},  \\ \label{ex3}
E _4 (x) = F[\exp ( - qx) \pm \exp (qx)]\exp(i\beta z),  \qquad \mbox{in region 4},  \label{ex4}
\end{eqnarray}
where $h=\sqrt{n_1^2k_0^2-\beta^2}$, $p=\sqrt{\beta^2-n_0^2k_0^2}$
, $q=\sqrt{\beta^2-n_2^2k_0^2}$ ($\beta$ and $k_0$ are the
propagation constant along the $z$ direction and the wave number
in vacuum, respectively), and $A,B,C,D,E$ and $F$ are the
coefficients to be determined. The $``+"$ and $``-"$ signs in Eq.
(\ref{ex4}) correspond to the symmetric supermode and the
antisymmetric supermode, respectively. From the continuity of the
tangential electric and magnetic fields at all horizontal
boundaries, we obtain the following characteristic equation for
the propagation constant $\beta$ of a supermode
\begin{eqnarray}
 \left( {\frac{{h\cos(h d_1)  + p\sin(h d_1) }}{h}} \right)\left [\frac{{q\mu _c e^{ - p(d_0-d_2)/2 } }}{{2\mu _2 }}(\sigma e^{qd_2/2 }  -
 e^{ - qd_2/2 } ) - \frac{{pe^{ - p(d_0-d_2)/2 } }}{2}(\sigma e^{qd_2/2 }  + e^{ - qd_2/2 } ) + \right. \nonumber \\
\left. \frac{{q\mu _c e^{p(d_0-d_2)/2 } }}{{2\mu _2 }}(\sigma e^{qd_2/2 }  - e^{ - qd_2/2 } ) + \frac{{pe^{p(d_0-d_2)/2 } }}{2}(\sigma e^{qd_2/2 }  + e^{ - qd_2/2 } ) \right ] -  \nonumber \\
 (h\sin(h d_1)  - p\cos(h d_1) )\left [\frac{{q\mu _c e^{ - p(d_0-d_2)/2 } }}{{2\mu _2 p}}( - \sigma e^{qd_2/2 }  + e^{ - qd_2/2 } ) + \frac{{e^{ - p(d_0-d_2)/2 } }}{2}(\sigma e^{qd_2/2 }  + e^{ - qd_2/2 } ) + \right . \nonumber \\
\left . \frac{{q\mu _c e^{p(d_0-d_2)/2 } }}{{2\mu _2 p}}(\sigma e^{qd_2/2 }  - e^{ - qd_2/2 } ) + \frac{{e^{p(d_0-d_2)/2 } }}{2}(\sigma e^{qd_2/2 }  + e^{ - qd_2/2 } )\right ] = 0 ,
\label{cha}
\end{eqnarray}
where $\sigma=1$ (or $-1$) for the symmetric (or antisymmetric)
supermode and $\epsilon_c \mu_c=n_0^2$ ( $\epsilon_c$ and $\mu_c$
are the permittivity and permeability of the cladding,
respectively). Compared with the corresponding analysis for a
conventional multi-layered RHM structure, the only difference is
in the continuity of the magnetic field on the boundary between
the $3rd$ layer and the $4th$ layer (the LHM layer; $\mu_2$ is
negative in the $4th$ layer while $u_c$ is positive in the $3rd$
layer). After the propagation constants $\beta_s$ and $\beta_a$
(for symmetric and antisymmetric supermodes, respectively) are
determined from the above equation, the coupling lengths (within
which the  light is coupled totally from one SM RHM waveguide to
the other) for this coupling system   can then be calculated by
\begin{eqnarray}
L=\pi /(\beta_s-\beta_a).
\label{length}
\end{eqnarray}

\section{NUMERICAL RESULTS AND DISCUSSION}

\begin{figure}
\includegraphics[width=5in]{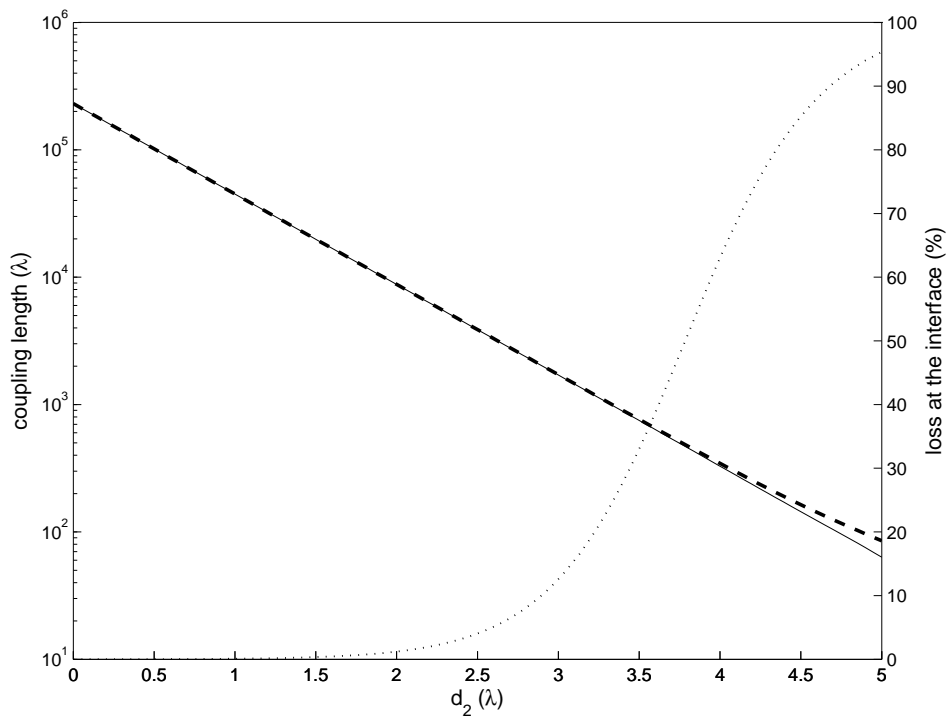}
\caption{\label{len1}The coupling length (solid line) and the
reflection loss (dots; at one end of the LHM layer) as the width
of the LHM layer increases when the ends of the LHM layer are not
tapered. The dashed line shows the total coupling length when the
LHM layer is tapered slowly at the two ends (the tapering angle is
fixed to $tg\theta=1/5$; the reflection loss is negligible for
this case).}
\end{figure}

In our simulations, the (effective) refractive indices of the core
and cladding are assumed to be $n_1=1.47$ and $n_0=1.46$,
respectively. We choose the width for each SM RHM waveguide as
$d_1=0.82d_s$ (where
$d_s=\lambda/(2\sqrt{n_1^2-n_0^2})=2.921\lambda$ is the cutoff
width) to keep the waveguide single-moded for a specific
wavelength $\lambda $, and we fix the separation between the two
SM waveguides as $d_0=4d_1$ (i.e., $9.58 \lambda$). Before
inserting the LHM layer (i.e., $d_2 = 0$), Eq. (\ref{cha}) gives a
coupling length $L=2.30\times10^5 \lambda$, which is about 35
centimeter if we choose a typical wavelength $\lambda =1.55 \mu m$
for optical communications. Obviously, such a directional coupler
is useless in any practical application as the coupling length is
too long. From the physical mechanism of the coupling, we see that
the coupling can be enhanced effectively (and thus the coupling
length can be shortened significantly) if we enlarge the field
amplitude of the evanescent wave  (of a SM RHM waveguide)
penetrating into the core of the other SM RHM waveguide. It is
well known that a LHM can amplify evanescent waves (see e.g.
\cite{pen2}). Thus,  we insert a LHM layer between the two SM RHM
waveguides (as shown in Fig. \ref{coupler}(a)) to enhance the
coupling. In our numerical simulation, we choose $\epsilon_2= -
\epsilon_c$ and $\mu_2=- \mu_c$ (i.e., the impedance and
refractive index for the LHM layer are matched with the ones for
the cladding). The coupling length calculated with Eq.
(\ref{length}) is shown by the solid line in Fig. \ref{len1} as
the width $d_2$ of the LHM layer increases. From this figure one
sees that the coupling length is shortened dramatically when the
width of the LHM layer  increases. Due to its amplification
ability for evanescent waves, the LHM layer enhances effectively
the coupling between the two SM RHM waveguides. When the width of
the LHM layer increases from $0$ to $5  \lambda $, the coupling
length is shortened rapidly from $L=2.30\times10^5 \lambda$ to $L=
63.20\lambda$ (less than 100 $\mu m$).

Now we consider the reflection loss at the start ($z=0$) of the LHM layer, which has a finite length (see Fig. 1(a); the reflection loss at the other end of the LHM layer has the same value).
Let an incident wave impinge on one of the SM RHM waveguides and propagate
along the $z$ direction into our directional coupler.
When it meets the start of the LHM layer at $z=0$,
the mismatch between the supermodes at $z= 0^-$ and $z= 0^+$ will cause
a reflection loss at the interface $z=0$.
This reflection loss can be estimated from the orthogonality of eigenmodes and an overlap integral method \cite{overlap}.
The dotted line in Fig. \ref{len1} shows the reflection loss (in terms of the energy) around the interface
$z=0$ as the width of the LHM layer increases.
As expected, when there is no LHM layer (i.e., $d_2 =0$), the reflection loss at $z=0$ is 0.
As the width of the LHM layer increases,
the reflection loss starts to appear, however,  keeps at a low level until $d_2$ reaches $ 2\lambda$
(note that the coupling length is shortened rapidly as the width of the LHM layer increases).
The reflection loss starts to increase quickly with the width $d_2$ after it reaches $ 2\lambda$. For example, about $62\%$ of the incident energy will be reflected around the interface
$z=0$ when $d_2=4\lambda$.

To reduce the reflection loss at the two ends of the LHM layer, we use a LHM layer with two slowly tapered ends
(see Fig. 1(b)).
Almost all the energy can then be coupled into the system (i.e., the reflection loss is negligible) when the
tapering angle $\theta$ is small enough.  In our simulation, we fix $tg\theta=1/5$.
For such a directional coupler, the propagation constants of the two supermodes vary slowly along the $z$ direction and the central strong coupling length $L_c$ should be determined from the following formula
\begin{eqnarray}
\int_0^{d_2 ctg\theta + L_c} {\beta _s dz - \int_0^{ d_2 ctg\theta + L_c} {\beta _a dz = \phi } } ,
\label{length2}
\end{eqnarray}
where the total phase difference $\phi $ is determined by the
desired  splitting ratio of the directional coupler (note that the
normalized output powers at the two SM RHM waveguides are $(1 +
\cos \phi)/2$   and $(1 - \cos \phi)/2$, respectively). When $
\phi =  \pi $,  the coupling length $(d_2 \, ctg\theta + L_c)$
gives the total length of the LHM layer (including the two tapered
parts) within which the light is coupled totally from one SM RHM
waveguide to the other. This total coupling length is shown by the
dashed line in Fig. 2 when the width $d_2$ of the LHM layer (at
the central part) increases. When the width of the LHM layer is
small, the coupling length is large and the tapering length is
vary small (relatively), and thus the difference between the
dashed line and the  solid line is small in Fig. 2. When the width
$d_2$ of the LHM layer becomes large, the coupling length is
shortened significantly and consequently the difference between
the coupling lengths for the two cases (with and without tapered
ends) becomes noticeable in Fig. 2. The tapering length at each
end is $25 \lambda$ when $d_2=5 \lambda$. For example, when we
increase the width $d_2$ of the LHM layer from $0$ to $5 \lambda
$, the total coupling length $( L= d_2 ctg\theta + L_c )$
decreases from $ 2.30\times10^5 \lambda$ to $ 85.11\lambda$. This
length ($ 85.11\lambda$) is about $34\%$ longer than the
corresponding coupling length ($L= 63.20\lambda$) when the LHM
layer has no tapered ends, but is still very short and thus the
device can be very compact. In the mean time, the tapering  makes
the reflection loss at the ends negligible  (and thus we do not
show it in Fig. 2).

\section{CONCLUSION}
A novel directional coupler formed by inserting a LHM layer of
finite length between two SM RHM waveguides has been introduced.
Using the method of supermodes, we have analyzed this coupling
system and studied how the coupling length for the two SM RHM
waveguides is shortened dramatically as the width of the LHM layer
increases. The coupling is enhanced through the amplification of
the evanescent waves in the LHM layer. The reflection loss due to
the sudden mismatch of supermodes at the two ends of the LHM layer
has also been estimated. To reduce this reflection loss, a LHM
layer with two slowly tapered ends has been used.  The novel
directional coupler can shorten the coupling length dramatically
without any bent waveguide.

The partial support of National
Natural Science Foundation of China under a key project (90101024) and a project (60277018) are
gratefully acknowledged.

\newpage

\end{document}